\documentclass[prl,floatfix,twocolumn]{revtex4}
\usepackage{amsfonts}
\usepackage{graphicx}

\def\be{\begin{eqnarray}}
\def\ee{\end{eqnarray}}
\def\el{\nonumber\\}
\def\tr{\mathrm{Tr}}
\def\dslash{\makebox[0cm][l]{$\,/$}D}
\def\1{\openone}
\def\e{\mathrm{e}}
\def\re{Re}
\def\conj#1{{{#1}^{*}}}
\def\zs{\eta}
\def\mus{\mu_s}
\def\ms{\hat{m}}

\begin{document}
\title{Universal results from an alternate random matrix model\\
 for QCD with a baryon chemical potential}
\author{James C. Osborn}
\affiliation{Physics Department, University of Utah,
Salt Lake City, UT 84112, USA}
\begin{abstract}
We introduce a new non-Hermitian random matrix model for QCD with
a baryon chemical potential.
This model is a direct chiral extension of a previously studied model
that interpolates between the Wigner-Dyson and Ginibre ensembles.
We present exact results for all eigenvalue correlations for any number
of quark flavors using the orthogonal polynomial method.
We also find that the parameters of the model can be scaled to remove
the effects of the chemical potential from all thermodynamic quantities
until the finite density phase transition is reached.
This makes the model and its extensions well suited for studying the 
phase diagram of QCD.
\end{abstract}
\maketitle

Random matrix models have long been used as effective models 
for a variety of complex quantum systems.
More recently the class of non-Hermitian random matrix models has
received much attention in systems including the fractional quantum
hall effect \cite{DGIL}, two dimensional charged plasmas \cite{FJ} and
quantum chaotic scattering (see \cite{FS} for a review).
Additionally one was introduced by Stephanov as a model for QCD
with a baryon chemical potential \cite{S}.
Since then this model and related extensions have often been used
to study the phase diagram of finite density QCD \cite{phase}.
It is also useful for evaluating numerical algorithms that might be used
to simulate finite density QCD \cite{numsim} and in studying the density
of the eigenvalues of the Dirac operator \cite{evcor}.
Despite the success of this model, exact results for the eigenvalue
correlations have remained notably absent.
Only recently has the eigenvalue density been obtained
in the low energy (microscopic) limit by Splittorff and Verbaarschot \cite{SV}.

Another model for the eigenvalues of the QCD Dirac operator with a chemical
potential was proposed by Akemann \cite{A1}.
This model was constructed as a chiral extension of a previously studied
non-Hermitian matrix model.
The main attraction of this model is that it can be solved exactly for
all the eigenvalue correlations.
However, the model is defined only in terms of the eigenvalues, and not
from a matrix model based on the symmetries of the Dirac operator with a
chemical potential.
It was later shown that the partition function for this model agreed with
that of Stephanov's in the microscopic limit \cite{A2}.
This agreement, however,  only holds for a sufficiently small
chemical potential as can be seen clearly by comparing the
microscopic spectral density \cite{SV}.

Here we will consider a new matrix model for QCD with a baryon chemical
potential which is also
based on the form of the Dirac operator like in Stephanov's model, but
can be solved for all spectral correlations like the model of Akemann.
We will give exact results for the correlations with any number of
quark flavors.
We will also consider the results in the microscopic limit where the
correlations should agree with QCD due to universality.
Lastly we look at the partition function and show that one can absorb the
effects of the chemical potential by scaling the parameters of the model.
This then provides a physically accurate model that shows no change
in thermodynamic quantities until the finite density phase transition is
reached.

The matrix model originally studied by Stephanov is 
\be
\label{dold}
\mathcal{D}_I ~=~ \left( \begin{array}{cc}
 0  & i A+\mu  \\
 i A^\dagger + \mu  & 0 
 \end{array} \right)
\ee
with $A$ a random complex matrix and $\mu$ is multiplied by the identity.
This is based on the form of the Dirac operator $\dslash + \mu \gamma_0$
in a chiral basis.
Here the matrix elements of $\gamma_0$ are chosen to be constant which
is accurate when $\mu$ is very large, but it makes
the model difficult to solve.

Here we consider an alternate basis where the matrix elements of
$\gamma_0$ are not taken as constant but are modeled with random matrices
as
\be
\label{dnew}
\mathcal{D}_{II} = \left( \begin{array}{cc}
0 & i A + \mu B \\
i A^{\dagger} + \mu B^{\dagger} & 0
\end{array} \right) ~.
\ee
Here $A$ and $B$ are complex $(N+\nu) \times N$ matrices
with $\nu$ the topological charge.
The QCD partition function with $N_f$ quark flavors can now be modeled as
\be
\label{pfnew}
Z = \mathcal{N}
 \int dA ~ dB ~ w_G(A) ~ w_G(B) ~
 \prod_{j=1}^{N_f} \det(\,\mathcal{D}_{II} + m_j\,)
\ee
where $\mathcal{N}$ is a normalization constant we will chose later.
At low energies the results are universal and do not depend on the
choice of the distribution for the matrix elements.
We therefore make the simple choice of a Gaussian measure given by
\be
\label{wg}
w_G(A) ~=~ \exp( \, - \, \alpha \, N \, \tr \, A^{\dagger} A \, ) ~.
\ee
At $\mu=0$ the parameter $\sqrt{\alpha} = \Sigma V / 2N$ is determined by the
absolute value of the chiral condensate in the chiral limit ($\Sigma$)
and the space-time volume ($V$).
We will consider the $\mu$ dependence of $\alpha$ later.
For now we will treat it as a constant and 
calculate the distribution of eigenvalues.

In order to diagonalize the Dirac matrix we first make the substitutions
$C = i A + \mu B$ and $D = i A^{\dagger} + \mu B^{\dagger}$.
The partition function becomes
\be
Z =
\mathcal{N}^{\prime}
\int dC \, dD \, w_{2}(C,D) \,
 \prod_{j=1}^{N_f}
 \det\left( \begin{array}{cc} m_j & C \\ D & m_j \end{array} \right)
\ee
with $\mathcal{N}^{\prime}$ a $\mu$ dependent constant and the weight is
\be
w_{2}(C,D) = 
\exp \bigg[
 &\!\!-&\!\!\!
 \frac{\alpha N (1+\mu^2)}{4 \mu^2} ~\tr(C^{\dagger} C + D^{\dagger} D) \el
 &\!\!-&\!\!\!
 \frac{\alpha N (1-\mu^2)}{4\mu^2} ~\tr(C D + C^{\dagger} D^{\dagger})
~\bigg] . ~~~~
\ee
The matrix $\mathcal{D}_{II}$ has exactly $\nu$ zero eigenvalues and
$N$ pairs of eigenvalues which we write as $\pm i z_k$.
The non-zero eigenvalues of $C D$ and $D C$ are then equal to $-z_k^2$.
We now introduce the parameterization
\be
C &=& U ( X + R ) V \nonumber\\
D &=& V^\dagger ( Y + S ) U^\dagger
\ee
where $U$ and $V$ are unitary square matrices,
$X$ and $R$ are complex $(N+\nu) \times N$
and $Y$ and $S$ complex $N \times (N+\nu)$ matrices.
The elements $X_{ij}$ and $Y_{ij}$ are zero except when $i=j$
and $R_{ij}$ and $S_{ij}$ are zero except when $i<j$.
The above decomposition is not unique and we must restrict the integration
measure on $U$ to the group U($N+\nu$)/[U(1)$^N$ $\otimes$ U($\nu$)] and
on $V$ to U($N$)/U(1)$^N$.
The Jacobian for this transformation can be calculated
\be
\left| \Delta(z^2) \right|^2 ~ \prod_k |x_k|^{2 \nu}
\ee
where $\Delta(z^2) = \prod_{i<j} (z_i^2-z_j^2)$
 is the Vandermonde determinant.
Here we have used that $x_k y_k = -z_k^2$ with $x_k$ and $y_k$ the diagonal
elements of $X$ and $Y$ respectively.
With this parameterization the integrals over $U$, $V$,
$R$ and $S$ become trivial and we are left with integrals over the
complex numbers $x_k$ and $y_k$.
We can change the integral over $y_k$ to an integral over the eigenvalues $z_k$
and then integrate out $x_k$ to arrive at the form
\be
\label{epfnew}
Z = \mathcal{N^{\prime\prime}} \prod_{j=1}^{N_f} m_j^{\nu}
 \int \left|\Delta(z^2)\right|^2 \,
 \prod_{k=1}^{N} w(z_k) \, dz_k
 \prod_{j=1}^{N_f} \, (z_k^2 + m_j^2) ~~
\ee
with $\mathcal{N^{\prime\prime}}$ another normalization constant and
\be
\label{wnew}
w(z) &=& |z|^{2\nu+2}
\exp\left( \frac{\alpha N (1-\mu^2)}{4 \mu^2} (z^2 + \conj{z}^2) \right) \el
&\times& K_\nu \left( \frac{\alpha N (1+\mu^2)}{2 \mu^2} |z|^2 \right)
\ee
where $K_\nu$ is a modified Bessel function.

The eigenvalue representation of the partition function (\ref{epfnew})
can be related to the model considered previously by Akemann.
For large argument the Bessel function becomes
$K_\nu(x) \approx \sqrt{\pi/2x}\exp(-x)$
which gives
\be
\label{wold}
w(z) &\approx& \sqrt{\frac{\pi \mu^2}{\alpha N (1+\mu^2)}} ~ |z|^{2\nu+1} \\
&\times&
\exp\left( - \frac{\alpha N (1+\mu^2)}{2 \mu^2}
 \left[|z|^2 - \frac{\tau}{2} (z^2 + \conj{z}^2) \right]\right)
\nonumber
\ee
where we have set $\tau = (1-\mu^2)/(1+\mu^2)$.
This is the same model considered in \cite{A1}.
The two models agree when
\be
\alpha N (1+\mu^2) |z_{min}|^2 \gg 2\mu^2 ~.
\ee
The smallest eigenvalue $z_{min}$ has an average size around
$\pi/(\Sigma V)$ at $\mu=0$.
It is then convenient to consider the above condition
in the microscopic \cite{ShV} and weak non-Hermiticity \cite{FKS} limits 
given by taking $N \to \infty$ while holding $\zs_k = \Sigma V z_k$,
$\mus^2 = 2N\mu^2$ and $\alpha$ fixed.
The above condition then becomes
\be
|\zs_{min}|^2 \gg 4 \mus^2 ~.
\ee
In the microscopic limit the two models agree only for sufficiently small
$\mu_s$.
Since the new model is based directly on the Dirac operator
we expect that (\ref{wnew}) is the correct eigenvalue weight for QCD
at low energies even for larger values of $\mu_s$.

\vspace{-3mm}
\section{Quenched spectrum}
\vspace{-3mm}

Given the expression for the partition function in terms of the
eigenvalues of the Dirac operator we can now solve for their correlations.
This was done in \cite{A1} for the model with weight (\ref{wold})
at $N_f = 0$ using the method of orthogonal
polynomials extended to the complex plane.
There it was shown that the corresponding orthogonal polynomials are the
Laguerre polynomials.
From that all spectral correlations can be determined analytically.
It turns out that the orthogonal polynomials corresponding to the weight
(\ref{wnew}) are also Laguerre.
We can therefore proceed in similar fashion.

One can easily verify that the monic polynomials
\be
\label{qp}
p_k^0 (z) = 
\left( \frac{\mu^2-1}{\alpha N} \right)^{k} k! ~
L_k^\nu\left(\frac{\alpha N z^2}{1-\mu^2}\right)
\ee
are orthogonal with respect to the integral
\be
\mathop{\int}_{\re(z)>0}
 p_k^0(z) ~ p_\ell^0(\conj{z}) ~ w(z) ~ dz 
~=~ h_k^0 ~ \delta_{k\ell}
\ee
with
\be
h_k^0 ~=~
\frac{\pi \, \mu^2 ~ (1+\mu^2)^{2k+\nu} ~ k! ~ (k+\nu)!}
     {2 \, (\alpha N)^{2k + \nu + 2}}  ~.
\ee
Here we restricted the range of integration to the positive real
half of the complex plane.
This avoids double counting pairs of eigenvalues with opposite signs
and ensures agreement with previously obtained results at $\mu=0$.

The $n$-point spectral correlator can be written as
\be
\label{rhoN0}
\rho_N^0(z_1, \ldots, z_n) = \det_{1 \le i,j \le n}
 \left[ \mathcal{K}_N^0(z_i, z_j) \right]
\ee
in terms of the kernel
\be
\mathcal{K}_N^0(x,y) = \sqrt{w(x) \, w(y)} \,
 \sum_{k=0}^{N-1} \frac{1}{h_k^0} \, p_k^0(x) \, p_k^0(\conj{y}) ~.
\ee
We can also evaluate the result in the microscopic limit.
The microscopic limit of the $n$-point correlator is
\be
\rho_s^0(\zs_1, \ldots, \zs_n) &=&
 \lim_{N \to \infty} \frac{1}{(\Sigma V)^{2n}}
 \rho_N \left(\frac{\zs_1}{\Sigma V}, \ldots, \frac{\zs_n}{\Sigma V} \right)~~~
\ee
with $\alpha$ kept fixed.
This limit can be taken following the procedure in \cite{A1}.
The result can be written as a determinant similar to (\ref{rhoN0})
but with the microscopic kernel 
\be
\label{k0m}
\mathcal{K}_s^0(x,y) &=& 
\frac{|xy|^{\nu+1}}{4 \pi \mus^2 (x\conj{y})^\nu}
\, \sqrt{
K_\nu\left(\frac{|x|^2}{4\mus^2}\right)
K_\nu\left(\frac{|y|^2}{4\mus^2}\right)
} \\
&\times& \e^{\frac{\re(x^2+y^2)}{8\mus^2}}
\int_0^1 \e^{-2\mus^2 t} J_\nu(x\sqrt{t}) J_\nu(\conj{y}\sqrt{t}) dt~~~
\nonumber
\ee
The spectral density is obtained by setting $x=y$ in the above expression
and agrees with the result given in \cite{SV}
based on the model using (\ref{dold}).
This agreement is expected since both models are based directly on the
form of the Dirac operator.
Only the microscopic results are universal, though, and the macroscopic
correlations of the two models do not necessarily agree.

\vspace{-3mm}
\section{Unquenched spectrum}
\vspace{-3mm}

We can also extend the results to an arbitrary number of flavors using
an iterative procedure similar to the one used for the case of $\mu=0$
in \cite{DN}.
However since the mass term is only a function of $z$ and not $\conj{z}$
we can't use a single set of orthogonal polynomials.
Instead we construct a set of biorthogonal polynomials for $a$ flavors
that satisfy the biorthogonality condition
\be
\mathop{\int}_{\re(z)>0}
 p_k^{a}(z) \, q_l^{a}(\conj{z})
 \, w(z) \prod_{j=1}^{a} \left(z^2+m_j^2\right) dz
= h_k^{a} \, \delta_{kl} .~~
\ee
The monic polynomials can be defined iteratively by
\be
\label{op}
p_k^{a}(z) &=& \frac
{p_{k+1}^{a-1}(z) \, p_k^{a-1}(i m_{a}) - p_k^{a-1}(z)\,p_{k+1}^{a-1}(i m_{a})}
{p_{k}^{a-1}(i m_{a}) ~ (z^2+m_{a}^2)} \el
q_k^{a}(\conj{z}) &=& \frac{h_k^{a-1}}{p_k^{a-1}(i m_{a})}
\sum_{n=0}^{k}\,\frac{p_n^{a-1}(i m_{a})}{h_n^{a-1}} \, q_n^{a-1}(\conj{z})
\ee
with the convention that $q_n^0(\conj{z}) = p_n^0(\conj{z})$.
The normalization coefficient can also be derived recursively by
\be
\label{norm}
h_k^{a} = - \frac{p_{k+1}^{a-1}(i m_{a})}{p_k^{a-1}(i m_{a})} \; h_k^{a-1} ~.
\ee
An explicit formula for the polynomials for a general weight function
appears in \cite{B}.

For convenience we will define the function
\be
\label{phidef}
\phi_k^a(z) = p_k^a(z) \prod_{j=1}^a (z^2+m_j^2) ~.
\ee
The $a$ flavor kernel can now be written as
\be
\mathcal{K}_N^a(x,y) =
 \tilde{\mathcal{K}}_N^a(x,y) \;
 \sqrt{w(x) \, w(y)} \,
 \prod_{j=1}^a \sqrt{\frac{y^2+m_j^2}{x^2+m_j^2}}
\ee
with
\be
\label{kt}
\tilde{\mathcal{K}}_N^a(x,y) = 
\sum_{k=0}^{N-1} \frac{1}{h_k^a} \; \phi_k^a(x) \; q_k^a(\conj{y}) ~.
\ee
We can get a recursion relation for this by substituting
(\ref{op}), (\ref{norm}) and (\ref{phidef}) into (\ref{kt}) and
rearranging the sums as
\be
\tilde{\mathcal{K}}_N^{a}(x,y) =
\tilde{\mathcal{K}}_N^{a-1}(x,y) -
\frac{\phi_N^{a-1}(x)}
     {\phi_N^{a-1}(i m_a)}
 \; \tilde{\mathcal{K}}_N^{a-1}(i m_a,y) .~~
\ee
Repeated application of these relations gives an expression for the
$a$ flavor kernel in terms of the zero flavor kernel and the functions
$p_{N+k}^0(x)$ for $0 \le k \le a-1$.
Notice that the modified kernel $\tilde{\mathcal{K}}_N^{a}(x,y)$ vanishes
at $x=i m_j$ for $0 \le j \le a$.
It can therefore be written in the form of a determinant
\be
\label{kernn}
\tilde{\mathcal{K}}_N^{a}(x,y) =
\frac{
\begin{array}{|cccc|}
\tilde{\mathcal{K}}_N^{0}(x,y)     & p_N^{0}(x) &
 \cdots & p_{N+a-1}^{0}(x) \\
\tilde{\mathcal{K}}_N^{0}(i m_1,y) & p_N^{0}(i m_1) &
 \cdots & p_{N+a-1}^{0}(i m_1) \\
\vdots & \vdots & \ddots & \vdots \\
\tilde{\mathcal{K}}_N^{0}(i m_a,y) & p_N^{0}(i m_a) &
 \cdots & p_{N+a-1}^{0}(i m_a)
\end{array}
}{\left|p_{N+\ell-1}^{0}(i m_{k})\right|_{1 \le k,\ell \le a}} .\el
\ee
The denominator was determined by requiring that the coefficient of
$\tilde{\mathcal{K}}_N^{0}(x,y)$ is unity.

We can also take the microscopic limit of the above kernel.
For this we also scale the masses with the inverse of the volume
keeping $\ms_k = \Sigma V m_k$ fixed.
This limit must be taken carefully and can be done with a procedure similar
to that in \cite{WGW}.
The result is
\be
\mathcal{K}_s^a(x,y) =
 \mathcal{K}_s^0(x,y)
 ~ \frac{\tilde{\mathcal{K}}_s^a(x,y)}{\tilde{\mathcal{K}}_s^0(x,y)}
 ~ \prod_{j=1}^a \sqrt{\frac{y^2+\ms_j^2}{x^2+\ms_j^2}}
\ee
with $\tilde{\mathcal{K}}_s^{a}(x,y)$ equal to
\be
\frac{
\begin{array}{|cccc|}
\tilde{\mathcal{K}}_s^0(x,y)     & J_{\nu}(x) &
 \cdots & x^{a-1}J_{\nu+a-1}(x) \\
\tilde{\mathcal{K}}_s^0(i \ms_1,y) & J_{\nu}(i \ms_1) &
 \cdots & (i\ms_1)^{a-1}J_{\nu+a-1}(i \ms_1) \\
\vdots & \vdots & \ddots & \vdots \\
\tilde{\mathcal{K}}_s^0(i \ms_a,y) & J_{\nu}(i\ms_a) &
 \cdots & (i\ms_a)^{a-1}J_{\nu+a-1}(i\ms_a)
\end{array}
}{\left|(i\ms_k)^{\ell-1}J_{\nu+\ell-1}(i\ms_{k})\right|_{1 \le k,\ell \le a}} . ~
\ee
The function $\tilde{\mathcal{K}}_s^0(x,y)$ is defined as the integral that
appears in (\ref{k0m}).
This gives the general result for the microscopic spectral correlations of the
QCD Dirac operator with a baryon chemical potential.

\vspace{-3mm}
\section{Partition function}
\vspace{-3mm}

This model can also be used as an effective model for studying qualitative
properties of the finite density QCD phase diagram.
We can construct this model so that it has the correct physical
behavior below the finite density phase transition.
To see this we will look at the partition function which is simply given by
\be
Z_N^{N_f} =
 \mathcal{N}^{''} N! \prod_{j=1}^{N_f} m_j^\nu \prod_{k=0}^{N-1} h_k^{N_f} ~.
\ee
Using the recursion relations (\ref{norm}) and (\ref{op}) the
partition function can be written in terms of the
quenched polynomials (\ref{qp}).
The result is
\be
Z_N^{N_f} = \mathcal{N}_N^{N_f} \prod_{j=1}^{N_f} m_j^\nu \;
 \frac{\det_{1\le k,\ell \le N_f}\left[ p_{N+\ell-1}^0(i m_k) \right]}
 {\Delta(m^2)}~.
\ee
This result can also be obtained from the more general expression 
given in \cite{AV}.

Notice that the polynomials $p_k^0$ only depend on $\mu$ through the
combination $\alpha/(1-\mu^2)$.
If we had started with a $\mu$ dependent $\alpha = (1-\mu^2)\, \alpha_0$
and an appropriate normalization $\mathcal{N}$ then we could have made a
partition function that doesn't depend on $\mu$.
This means that all thermodynamic quantities would remain at their $\mu=0$
values for $\mu < 1$.
This agrees with the phenomenological expectation that the baryon number
density should remain at zero until a first order phase transition is
reached.
In this respect the new model is again an improvement over the previous one
(\ref{dold}) since that one shows an unphysical behavior in the baryon
density as the chemical potential is increased.
Another matrix model with improved thermodynamic properties was studied in
\cite{H}, however, that model has not been solved for the eigenvalues.

At $\mu \ge 1$ the above scaling is no longer valid since the Gaussian
integrals in the partition function would not converge.
However, one can show that for $\mu>1$ and $\alpha>0$ the chiral condensate
in the unquenched theory vanishes in the large $N$ limit.
The model therefore does have a first order phase transition at $\mu=1$.
Of course for large $\mu$ the Dirac operator should approach the deterministic
form $\mu \gamma_0$ given in (\ref{dold}).
Thus this new model cannot be expected to correctly describe the eigenvalue
correlations in the phase where chiral symmetry is restored.
It is currently then a phenomenological model for QCD at
$\mu < \mu_c$ (here $\mu_c = 1$).

Given this models success it would be interesting to include temperature
into the model and study the full phase diagram as has done with
previous matrix models \cite{phase}.
A more complete analysis of the phase diagram will be saved for a later
publication.
Another interesting extension would be to make both $A$ and $B$ be
random banded matrices with a power law decay like what was considered
for $\mu=0$ in \cite{GO}.
This was found to provide good agreement with an instanton liquid model for
eigenvalue correlations at larger energies (beyond the Thouless energy).
This also may provide an accurate way to study the finite temperature
chiral restoration transition.

In conclusion we have introduced a new random matrix model for
QCD with a baryon chemical potential that is based on the form of the
Dirac operator in a chiral basis
and is exactly solvable for all eigenvalue correlations.
We have presented the correlations for any number of quark flavors and
given results in the microscopic limit where they should agree with
QCD due to universality.
This new model also has a physically accurate thermodynamic behavior for
$\mu<\mu_c$ and could provide a basis for a phenomenological model of the
finite density QCD phase diagram.

I would like to thank G. Akemann and J.J.M. Verbaarschot for helpful
discussions.
This work was supported in part by NSF PHY 01-39929.

\end{document}